\documentclass[num-refs]{wiley-article}

\usepackage{siunitx}

\papertype{Original Article}
\paperfield{Journal Section}

\title{An Unconventional Ultra-Sub-Wavelength Receiving 
 Nano-Antenna Activated by ac Spin Pumping and the ac Inverse Spin Hall Effect}

\author[1]{Raisa Fabiha}
\author[1]{Michael Suche}
\author[1]{Erdem Topsakal}
\author[1]{Supriyo Bandyopadhyay}

\affil[1]{Department of Electrical and Computer Engineering, Virginia Commonwealth University, Richmond, VA 23284, USA}

\corraddress{Supriyo Bandyopadhyay, Department of Electrical and Computer Engineering, Virginia Commonwealth University, Richmond, VA 23284, USA}
\corremail{sbandy@vcu.edu}

\runningauthor{Fabiha, et al.}

\begin{document}

\begin{frontmatter}
\maketitle

\begin{abstract}
We report an extreme sub-wavelength unconventional receiving antenna. It consists of an array of nanomagnets connected to  heavy metal nanostrips. Incident electromagnetic (EM) radiation generates intrinsic and extrinsic spin waves in the nanomagnets, which pump spin into the heavy metal nanostrips at their own frequencies giving rise to a polychromatic alternating voltage across the latter owing to the ac inverse spin Hall effect. This implements a  receiving nano-antenna. We demonstrate its operation at two different EM wave frequencies of 1.5 GHz and 2.4 GHz -- the latter being the Bluetooth and Wi-Fi frequency. We measure the receiving gain at 2.4 GHz to be $\sim$ - 9 db. The free space radiated wavelength $\lambda$ at 2.4 GHz is 12.5 cm while the antenna area $A$ is merely 160 $\mu$m$^2$, making the ratio $A/\lambda^2$ = 0.97$\times$10$^{-8}$. This antenna's receiving gain should be very poor because of the tiny size. Yet the measured gain is more than  4,000 times larger than the theoretical limit for a conventional antenna of this size at this wavelength because of the unconventional operating principle.

\end{abstract}
\end{frontmatter}

\section{Introduction}
When electromagnetic (EM) radiation is incident on a ferromagnet, it excites spin waves in the latter. This can generally happen in two ways. First, the oscillating magnetic field in the EM radiation can excite spin waves  \cite{hillebrands,ganguly,serga}.
Second, parametric pumping can excite spin waves at half the frequency of the oscillating magnetic field \cite{serga, gurevich,lvov}. An additional consideration comes into play when the ferromagnet is patterned into {\it nanostructures} and arranged in a periodic two-dimensional array (sometimes referred to as a ``magnonic crystal''). In this case, very specific confined modes (such as center modes, edge modes, quantized modes, etc.) can be excited in the nanomagnets by the EM fields. They are of two types -- {\it intrinsic} and {\it extrinsic} \cite{nanoscale}. The intrinsic  mode frequencies are determined by such parameters as the shape and size of the nanomagnets \cite{suzuki}, as well as the pitch of the array, and may not have any relation to the EM wave frequency. The extrinsic mode frequency is the same as that of the excitation, namely the EM wave. Both intrinsic and extrinsic modes are spawned by transferring energy from the EM wave to the spin waves by photon-magnon coupling which has recently been shown to be quite strong in these systems \cite{salikhov}.

\begin{figure}[!h]
\vspace{-0.6in}
\centering
\includegraphics[angle=270,width=6in]{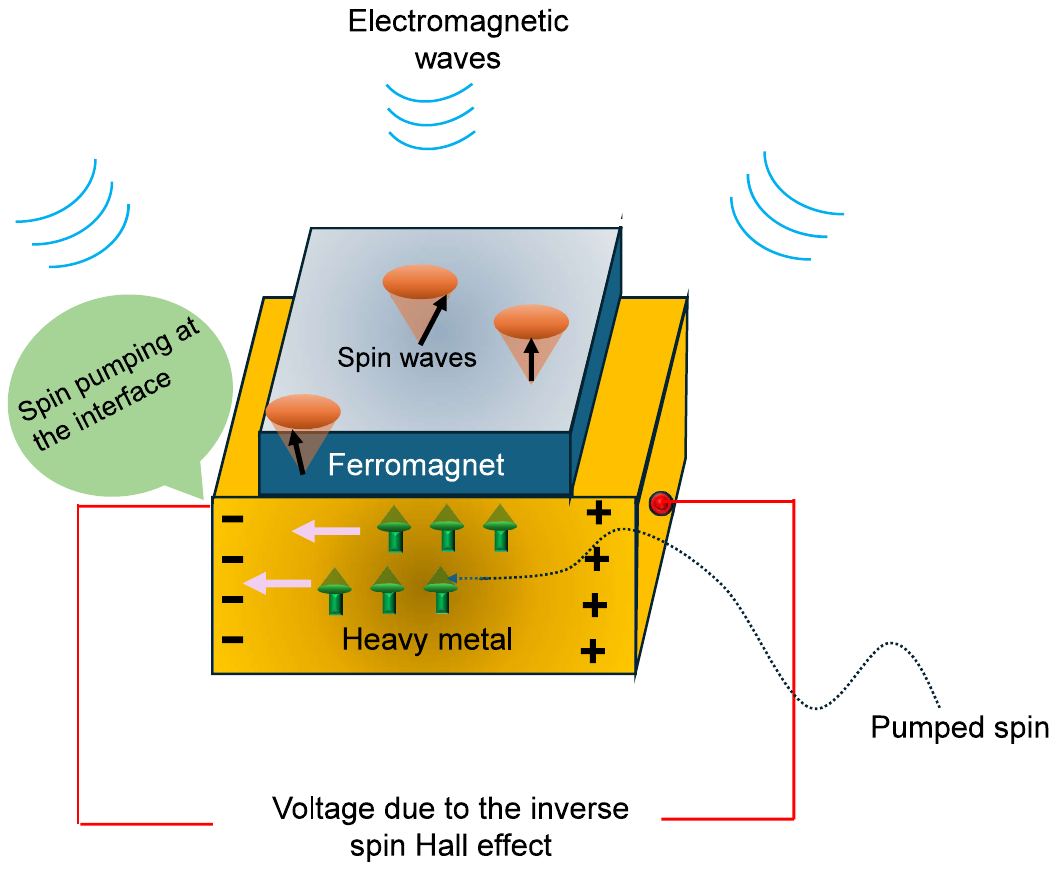}
\vspace{-0.8in}
\caption{Operating principle of the receiving antenna based on ac spin pumping and the ac inverse spin Hall effect. EM radiation excites spin waves in the ferromagnet which pumps spin into the heavy metal layer and that causes an ac voltage to appear across the latter which can be electrically detected to implement a receiving antenna.}
\label{fig:principle}
\end{figure}

If the nanomagnets are in physical contact with a heavy metal that exhibits the spin Hall effect (e.g. Pt), then the EM-excited intrinsic and extrinsic spin waves can {\it pump} spin into the heavy metal \cite{brataas} at their own frequencies and that can cause a polychromatic ac voltage to appear across the heavy metal via the ac inverse spin Hall effect \cite{brataas,silva,youssef,woltersdorf,jiao}. This voltage's frequency components will correspond to frequencies of the intrinsic and extrinsic modes that are excited in the nanomagnets by the EM field. The appearance of this voltage signals the presence of the EM radiation and hence implements a receiving antenna.

The excitation and amplification of intrinsic spin wave modes and generation of an extrinsic mode in a magnonic crystal by a surface acoustic wave (SAW)  was recently demonstrated by us \cite{nanoscale}. In a magnetostrictive nanomagnet, the SAW  effectively produces an alternating magnetic field directed in the direction of SAW propagation \cite{nanoscale}. The EM wave also generates an alternating magnetic field in the nanomagnet and therefore is similar in effect to a SAW. Consequently, it will also excite and/or amplify intrinsic modes and produce an extrinsic mode at the EM wave frequency in the nanomagnets. Both or either of the effects can be manifested. These  spin wave modes will pump spin into the heavy metal nanostrips that are in contact with the nanomagnets and cause an ac voltage to appear across the heavy metal owing to the inverse ac spin Hall effect. That realizes receiving antenna operation. The generated voltage will contain frequency components corresponding to the frequencies of the excited spin wave modes. The underlying principle of this voltage generation (i.e. the receiver antenna operation) is illustrated in Fig. \ref{fig:principle}. 

The converse effect leading to a {\it transmitting} antenna was recently demonstrated by us \cite{arxiv}. By injecting an alternating charge current into a heavy metal nanostrip in contact with an array of nanomagnets, we were able to excite extrinsic and intrinsic spin waves in the nanomagnets via the spin Hall effect. The spin waves then radiated EM waves into the surrounding medium by virtue of photon-magnon coupling \cite{arxiv}. That constituted a  ``transmitting'' antenna. Here, we use the same or similar samples to demonstrate the {\it receiving} functionality based on spin pumping and the inverse spin Hall effect. This demonstrates the complete transceiver system based on the spin Hall effect and the inverse spin Hall effect.

\section{Experiment}
\vspace{-0.1in}
\begin{figure}[!h]
\centering
\includegraphics[angle=270,width=6in]{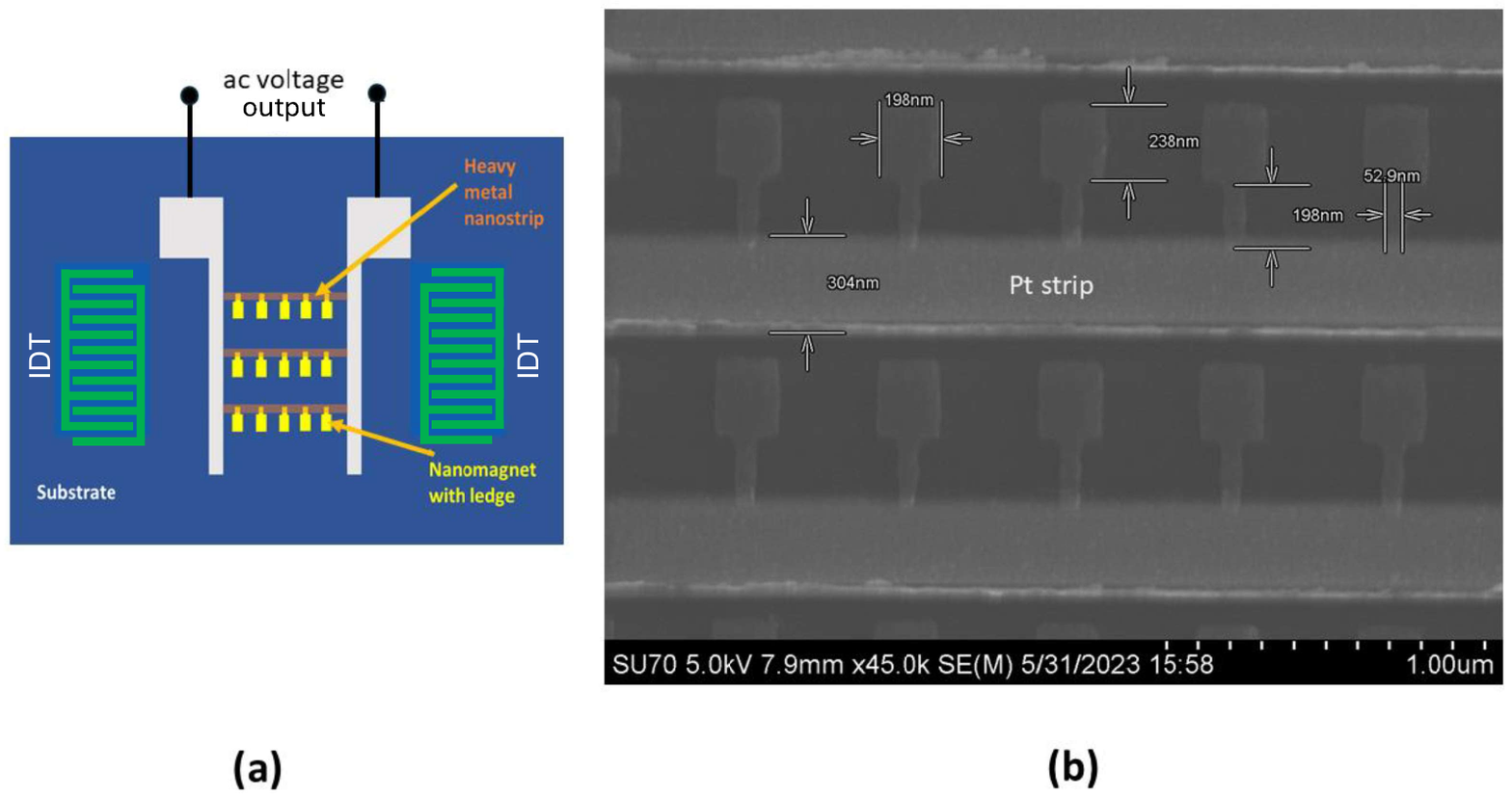}
\vspace{-0.6in}
\caption{(a) Schematic of the receiving antenna. There are parallel linear arrays of ledged nanomagnets (in yellow) with a Pt nanostrip (in orange) underlying each line. Each nanostrip connets to two common contact pads between which the received signal (ac voltage output) is detected on an oscilloscope. There are two interdigitated transducers (IDTs) on the two sides to pick up any surface acoustic waves produced by the EM radiation (not used in this experiment). (b) Scanning electron micrograph of the sample showing all relevant dimensions. This figure is adapted from \cite{arxiv}. }
\label{fig:receiver}
\end{figure}

The schematic of our receiving antenna device is shown in Fig. \ref{fig:receiver}(a). It consists of linear periodic arrays of ``ledged'' rectangular 15-nm thick cobalt nanomagnets deposited on a Si substrate, with a heavy metal (Pt) nanostrip underlying the ledges as shown in Fig. \ref{fig:receiver}(a).  The Pt nanostrip is $\sim$300 nm wide and 5 nm thick. There are 3000 linear arrays,  each containing 95 nanomagnets (total of 285,000 nanomagnets), and the ends of the nanostrips in each array are connected to two contact pads as shown in Fig. \ref{fig:receiver}(a). The total area covered by the nanomagnets and the Pt strip is about 160 $\mu$m$^2$. Fig. \ref{fig:receiver}(b) shows a scanning electron micrograph of the sample with all the relevant dimensions.  The inter-nanomagnet separation is large enough that any dipole interaction between neighbors is negligible.

\subsection{Why are the nanomagnets ledged?}

There is a reason why the nanomagnets are ledged and the Pt strip is placed only over the ledges. Since Co is magnetostrictive, it physically expands and contracts when its magnetization alternates with the frequency of the spin wave(s) generated by the EM radiation. If we place the Pt strip directly on the nanomagnets it will ``clamp'' the nanomagnets and prevent the expansion/contraction, which will quench the spin waves, spin pumping and  the ensuing detection process that implements the receiving antenna. It is imperative to {\it not} clamp the nanomagnets if they are magnetostrictive. This is the reason that the Pt strip is placed only on the ledges so as to not encumber the expansion/contraction of the bulk of the nanomagnets. Spin is pumped into the bulk of the nanomagnets {\it through} the ledges. The bottom surfaces of the nanomagnets are of course clamped by the underlying substrate, but this does not hinder the expansion/contraction of the top layers and hence does not prevent the excitation of intrinsic and extrinsic spin wave modes within the nanomagnets by the EM radiation.

This begs the question as to why not use a non-magnetostrictive material for the nanomagnets to eliminate this complication. The reason is that if the EM waves are generating spin waves in the nanomagnets and causing them to expand and contract periodically, then that would also cause alternating strain and a surface acoustic wave to propagate in the substrate, which can be picked by by interdigitated transducers (or solid electrodes) if the substrate were piezoelectric \cite{abeed}. This provides a second alternate way of detecting the EM radiation, making it a ``dual mode'' detector or receiving antenna. In this experiment, we did not use a piezoelectric substrate and the dual modality of detection.

\section{Testing the receiver function}

We fabricated two samples -- one with the nanomagnets and the other without. We call the former the ``real sample'' and the latter the ``control sample''. We compare the signals received from the two samples to eliminate spurious effects. If our theory is correct, the control sample will not generate any ac voltage output, but the real sample will.

\begin{figure}[!hbt]
\vspace{-0.8in}
\centering
\includegraphics[angle=270,width=6in]{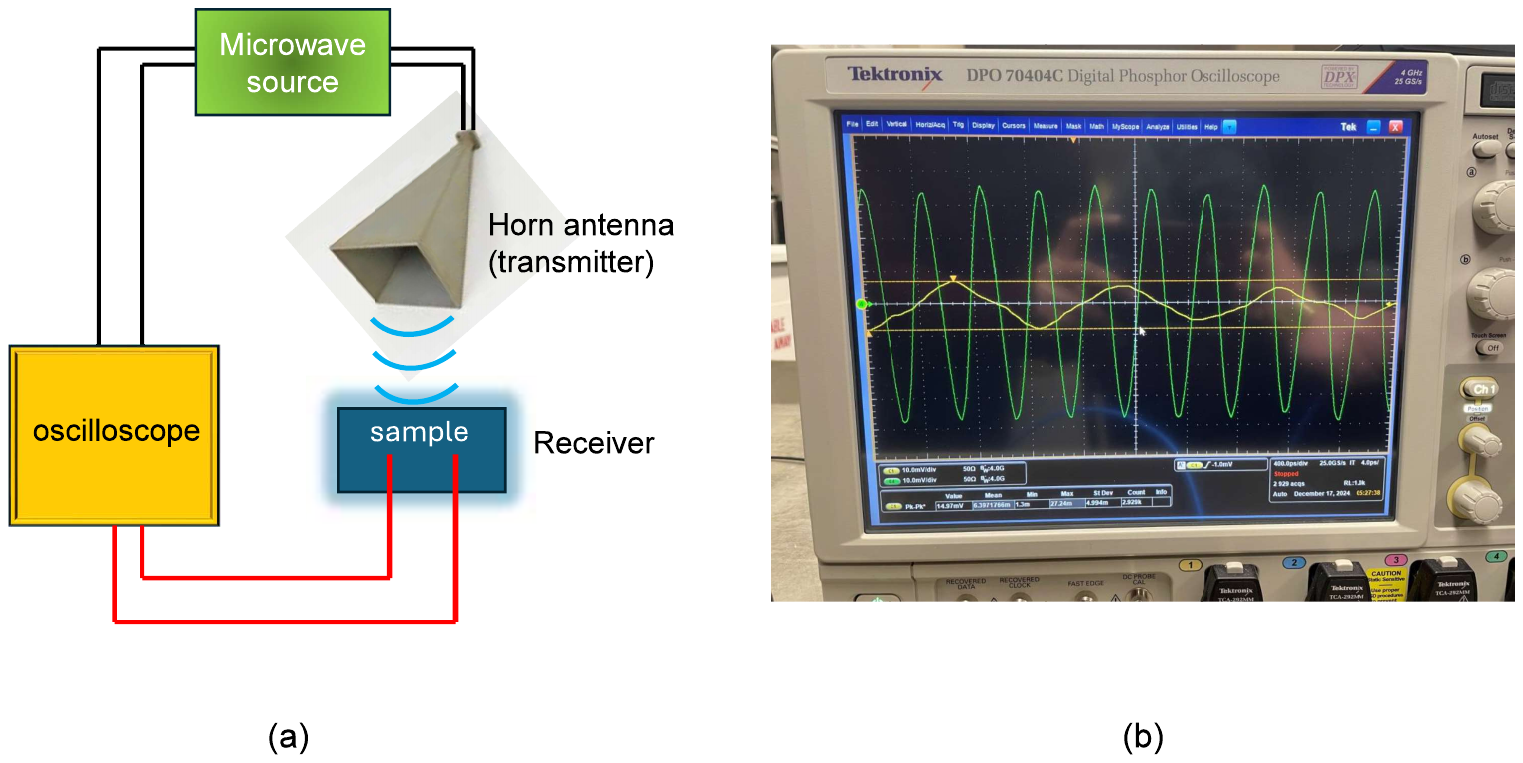}
\vspace{-0.6in}
\caption{(a) Schematic of the experimental set up. (b) Oscilloscope traces for the real sample when the separation between the horn antenna and the sample is 6 in. The green trace is the signal fed to the horn antenna (transmitted signal) and the yellow trace is the output measured between the two contact pads of the real sample (received signal). They are both on the same scale of 10 mV/div. The waveforms and periods are very different since the output has intrinsic and extrinsic modes mixed into it. In this case, the two signals are almost out of phase with each other and the time difference between a peak of the yellow trace and the closest peak of the green trace is 0.13 ns. The ratio of the peak-to-peak amplitudes of the two signals $V_{in}/V_{out}$ is $\sim$4:1.}
\label{fig:results}
\end{figure}

\begin{figure}[!hbt]
\vspace{-0.8in}
\centering
\includegraphics[angle=270,width=6in]{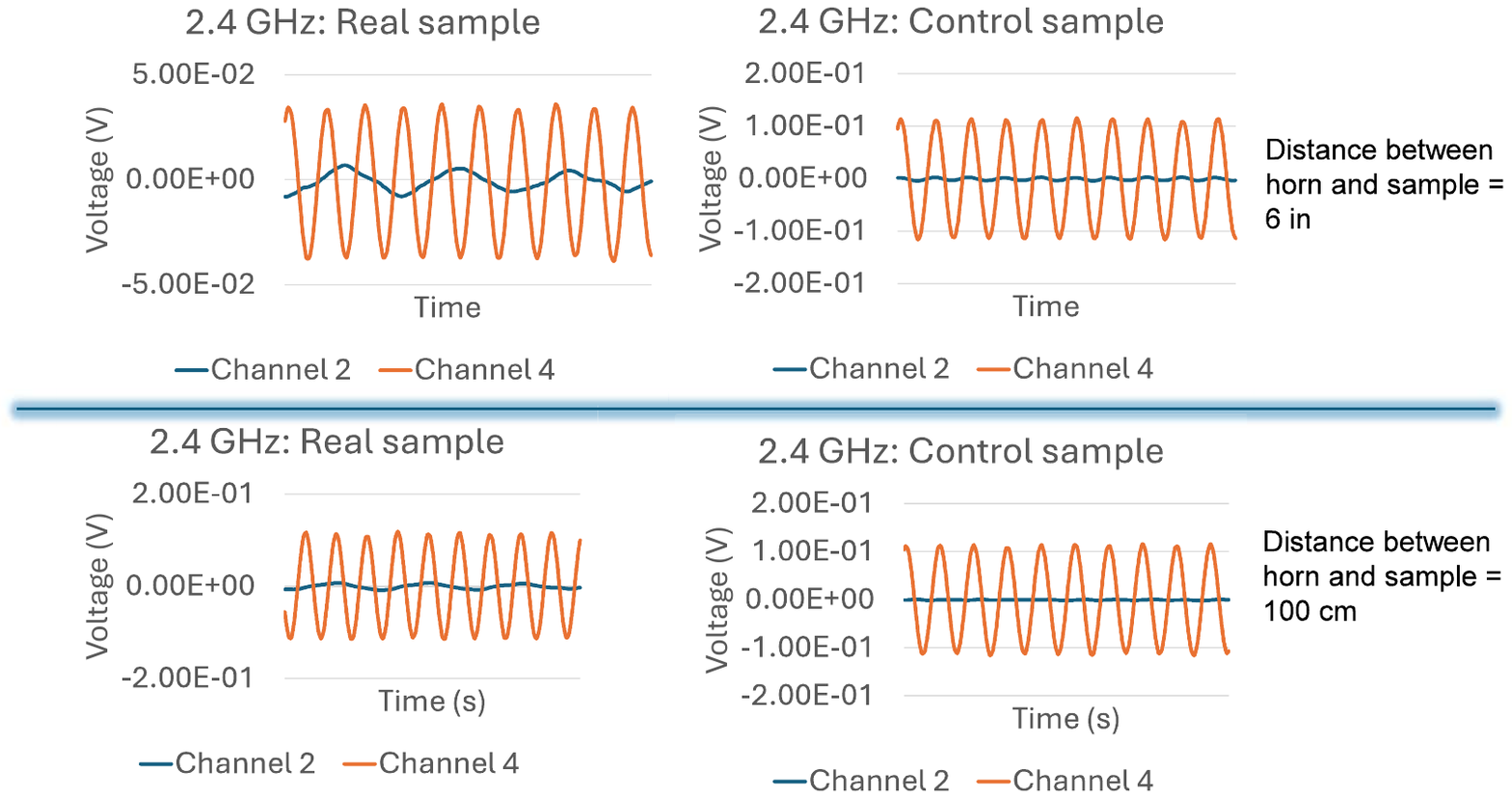}
\vspace{-0.6in}
\caption{Digitized oscilloscope traces of the input signals fed to the horn antenna (in orange) and the output signals produced at the two output contact pads (in blue). The upper panel corresponds to a horn-sample separation of 6 in and the lower panel to 100 cm. The input signal is fed at channel 2 of the oscilloscope and the output signal at channel 4. The left panel corresponds to the real sample and the right panel to the control sample.}
\label{fig:results1}
\end{figure}

Both samples were illuminated with a 2.4 GHz and a 1.5 GHz microwave signal transmitted with a horn antenna fed from a microwave source emitting 5 dbm of power as shown in Fig. \ref{fig:results}(a). The two output pads of the sample and the port of the microwave source were connected to two different channels of a microwave frequency digital  oscilloscope. The first channel displays the waveform of the transmitted signal emanating from the horn antenna and the second channel displays the waveform of the received signal measured between the two contact pads. We observed the oscilloscope traces in the two channels at two different separations between the sample and the transmitting horn antenna -- of 6 inches and 100 cm. 
The digital outputs of the oscilloscopes are plotted in  Fig. \ref{fig:results1}  for the two separations at 2.4 GHz excitation. The results for both the real sample and the control sample are shown.
We chose the frequency 2.4 GHz since it conforms to standard Bluetooth and Wi-Fi.

\section{Discussion}

Several features are observed in Fig. \ref{fig:results1}. Looking at the control sample results, we find that a small ac voltage appears between the output terminals when the distance between the horn and the sample is 6 inches. Nothing detectable appears when the distance increases to 100 cm (at the oscilloscope scale of 10 mV/div). More importantly, at 6 inches separation, the ac voltage detected at the control sample's output has the {\it same waveform} as the EM signal radiated by the horn antenna and there is virtually {\it no phase shift} between them. Therefore, this received signal in the control sample is most likely due to {\it electromagnetic pickup} through the air that has nothing to do with the receiver functionality. Signal from the horn antenna is traveling directly through air to the output contact pads of the control sample.

The story with the real sample is very different. There are many differences between the signal fed to the horn antenna (transmitted signal) and the signal appearing between the output contact pads (received signal): (1) they do not have the same frequency, (2) they do not have the same waveform and (3) there is a clear phase shift between the two. This eliminates electromagnetic pickup as the source of the output signal. It is the signal produced by transduction of EM waves to spin waves, followed by spin pumping, followed by the ac inverse spin Hall effect that causes the ouput voltage. The existence of this  signal establishes the detector (or receiving antenna) functionality.

\subsection{Frequency components of the received signal}

\begin{figure}[!ht]
\centering
\includegraphics[width=6in]{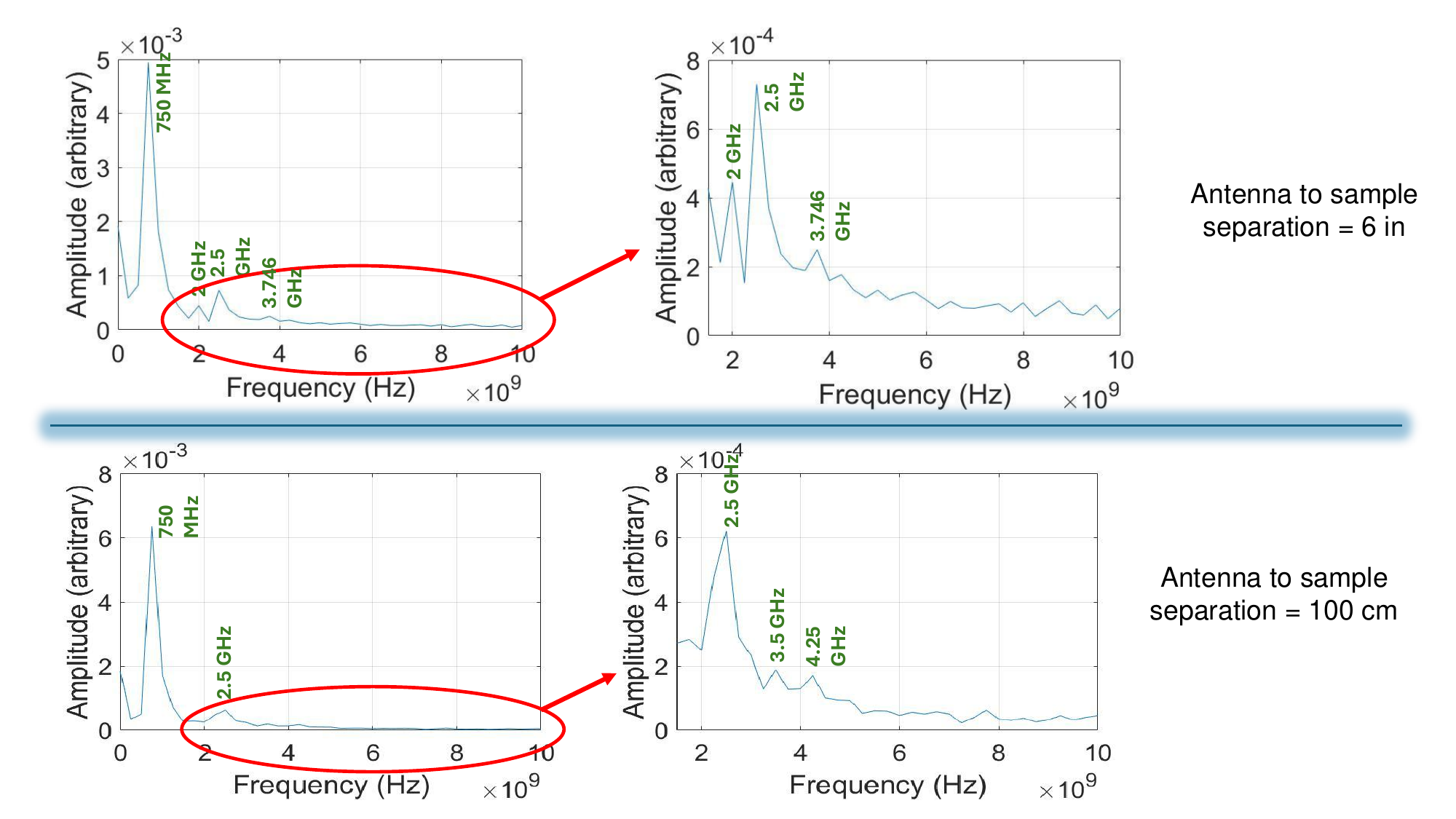}
\caption{Fast Fourier transform (FFT) of the received signal when the EM excitation frequency is 2.4 GHz. The upper panel shows the result when the horn-to-sample separation is 6 in and the lower panel shows the result when the separation is 100 cm. The left figure in either panel is the FFT showing all the peaks, whereas the right figure is a plot of the satellite peaks where the main peak at 750 MHz has been intentionally suppressed.}
\label{fig:Fourier}
\end{figure}

In Fig. \ref{fig:Fourier}, we show the fast Fourier transform of the received signal in  the real sample at 2.4 GHz excitation for the case when the horn-sample separation is 6 in and also for the case when the separation is 100 cm. Note that the dominant peak is at 750 MHz which is {\it not} the EM signal frequency of 2.4 GHz. There is also a satellite peak at 2.5 GHz. Both peaks show up independent of the horn-to-antenna separation, i.e., both at 6 in and 100 cm. There are other smaller peaks whose frequencies are not separation independent. 

The 750 MHz peak and the 2.5 GHz peak also have another feature. We show in the Supporting Information that these modes are present even when the excitation frequency is changed to 1.5 GHz from 2.4 GHz. Thus, they are independent of both separation and EM wave frequency. Because of these two features, we believe that they are intrinsic modes in the nanomagnet array. Their frequencies are determined by the size and shape of the nanomagnets \cite{suzuki} as well as other array parameters, and are independent of the EM frequency or separation between the transmitter and the receiver. 

Interestingly, at 2.4 GHz excitation, the EM wave does not spawn an extrinsic mode at its own frequency, but excites the intrinsic modes in the nanomagnet array by transferring energy to them. Nonetheless, these intrinsic modes also signal the presence of the EM field (they will be absent without the EM field) and hence fulfill the receiving antenna requirements. 

There are some separation-dependent modes at 2.0, 3.5, 3.75 and 4.25 GHz. These cannot be intrinsic modes because they are not independent of separation. They 
could be due to {\it vortex modes} caused by strain pulses generated in
the magnetostrictive nanomagnets by the EM radiation. The EM radiation very likely causes some periodic heating
and cooling of the nanomagnets, which, in turn, can cause periodic strain in the nanomagnets due the difference in
the thermal expansion coefficients of the nanomagnet material and the substrate. It has been shown that this can
spawn vortex modes in magnetostrictive nanomagnets \cite{cui}. If we change the separation between the transmiter and receiver, the heating rate will change and that will change the vortex modes and their frequencies. We emphasize that this is somewhat speculative but can explain the observed separation dependence.

Looking at the results for the control sample in the right panel of Fig. \ref{fig:results1}, we do not see any measurable signal produced between the contact pads at the horn-sample separation of 100 cm (when both oscilloscope channels have the same amplification of 10 mV/div) and the signal at the separation of 6 in clearly has only one frequency component (or at least a dominant frequency component) at 2.4 GHz which is the signal frequency. There is no frequency component at 750 MHz or 2.5 GHz since the control does not have any nanomagnet and hence no intrinsic spin wave modes that can form at 750 MHz or 2.5 GHz. The signal produced between the pads in the control sample is almost surely due to electromagnetic pick up because it is in phase with the input signal (within our measurement tolerance) and has the same frequency.

\subsection{Receiver gain}

We can calculate the receiver gain $G_r$ at 2.4 GHz from the formula \cite{Stutzman}
\begin{equation}
\frac{P_r}{P_t} \approx \frac{V_{out}^2}{V_{in}^2}  = G_t G_r {{\lambda^2 }\over{4 \pi R^2}},
    \end{equation}
where $\lambda$ is the wavelength, $R$ is the separation between the antenna and the sample, $P_r$ and $P_t$ are the received and transmitted power, respectively, $V_{out}$ and $V_{in}$ are the amplitudes of the waveform in the two oscilloscope channels, and $G_t$ is the gain of the transmitting ETS 3115 horn antenna at 2.4 GHz, which is 9.6 dB = 9.12, as provided by the manufacturer. This yields
\begin{equation*}
    G_r = 0.1277 = -8.9 db.
\end{equation*}

We can check if we get a very different gain from the 100 cm separation data. In this case $V{in}/V_{out} \approx 25$. Hence the gain is 
\begin{equation*}
    G_r = 0.116 = -9.3 db,   
\end{equation*}
which is very close to the previous value showing that the gain is roughly separation-independent at 2.4 GHz, and is approximately -9 db.

There are theoretical limits on the transmitting gains of conventional antennas that radiate via classical fluctuating electrical dipoles. Although they vary slightly depending on the exact type of the antenna, it is generally of the order \cite{Harrington,skrivervik}
\begin{equation}
    G_t^{max} = A/(2 \pi \lambda )^2 + \sqrt{A}/(\pi \lambda) ,
    \label{limit}
\end{equation}
where $A$ is the antenna area and $\lambda$ is the free space radiated wavelength. Because of the principle of reciprocity, we would expect the same limit to apply to the receiving gain. In our case, this limit turns out to be about 3.22$\times$10$^{-5}$ which is $\sim$ -45 db. 
Our gain is 36 db larger, i.e.,  about 4,000 times larger. However, this is not surprising for unconventional antennas that operate by unconventional means because in the past they have shown gains that exceed the theoretical limit in Equation (\ref{limit}) by more than two orders of magnitue \cite{raisa}. 

\section{Conclusion}
In conclusion, we have demonstrated a novel receiving antenna that can be aggressively miniaturized to dimensions that are orders of magnitude smaller than the free space radiated wavelength. Its gain exceeds the approximate theoretical limit for conventional antennas by more than three orders of magnitude. This is made possible by the unconventional modality of antenna activation whereby incident electromagnetic radiation excites spin waves in nanomagnets that pump spin into adjacent heavy metal nanostrips to generate alternating voltages across the latter owing to the inverse ac spin Hall effect.

The converse effect leading to a transmitting antenna was reported in \cite{arxiv} and it utilized an identical sample. Thus, both transmitter and receiver can be integrated on the same wafer with the same processing steps to produce a complete and novel (spintronic) transceiver system for on-chip communication.





\section*{Conflict of interest}
The authors declare no conflict of interest.





\pagebreak

\begin{center}
    {\Large \bf SUPPORTING INFORMATION}
\end{center}

\section*{Results for 1.5 GHz frequency}

We obtained results when the excitation frequency was 1.5 GHz instead of 2.4 GHz. These results are shown in Fig. \ref{fig:1.5} below.

\begin{figure}[!b]
\vspace{-0.2in}
\centering
\includegraphics[width=4in,angle=270]{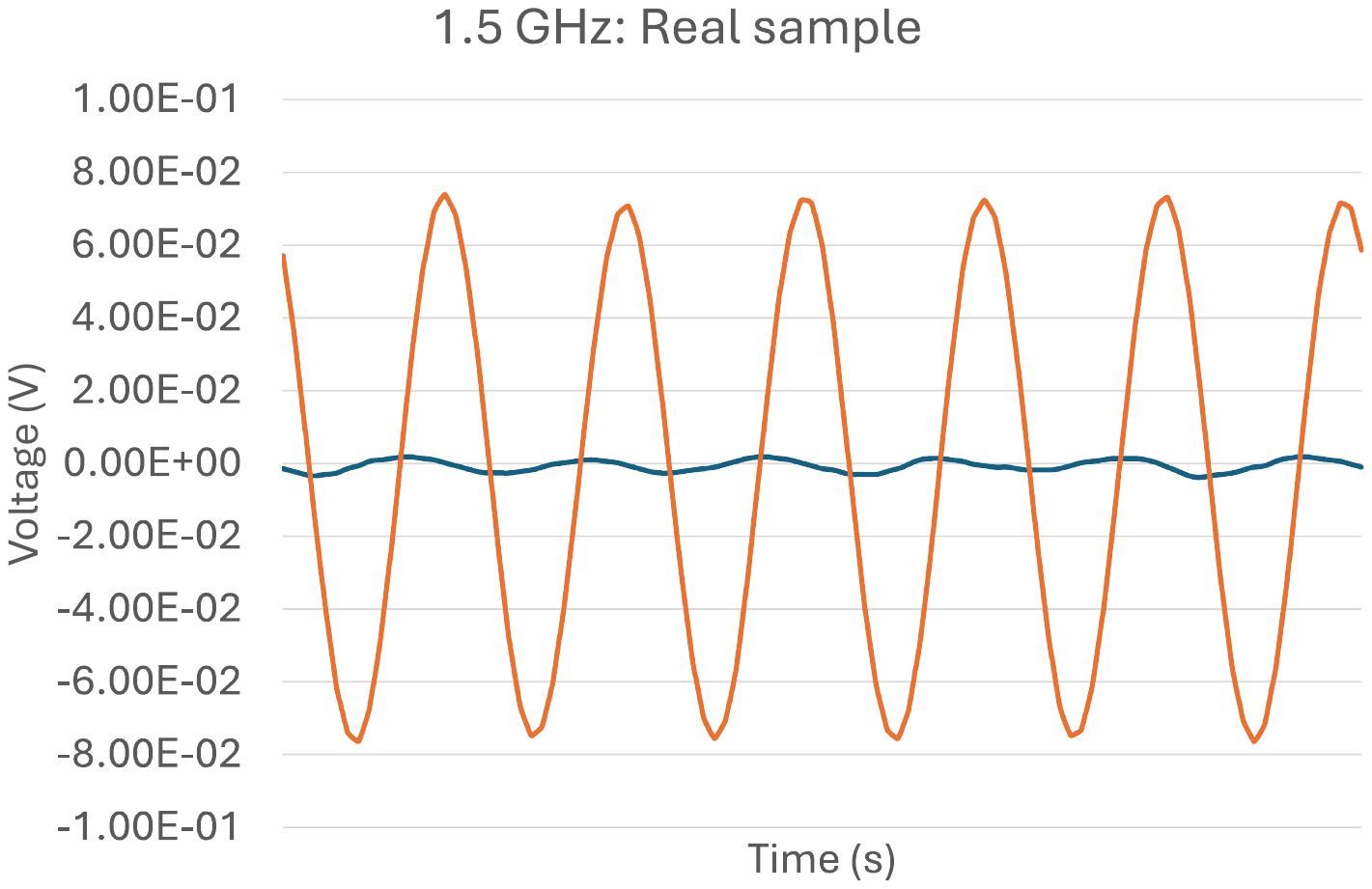}
\vspace{-0.6in}
\caption{Digitized oscilloscope traces of the input signal fed to the horn antenna (in orange) and the output
signal produced between the two output contact pads (in blue) in the real sample when the excitation frequency is 1.5 GHz and the horn-sample separation is 6 in. In this case, the amplitude of the input signal is roughly 15 times that of the output signal. There is a phase difference between the two which suggests that this is not due to direct electromagnetic pickup through the air.}
\label{fig:1.5}
\end{figure}

Unlike in the case of 2.4 GHz excitation, here the period and waveform of the input and output are the same, which raises the specter of the output signal being due to electromagnetic pickup. However, there is a phase difference between the two. The time taken by the electromagnetic wave to travel from the horn to the sample in this case will be $\Delta t$ = 6 in/3$\times$10$^8$ m/s = 0.5 ns and hence the phase shift between the input and output would have been $\phi = 2 \pi f \Delta t$ ($f$ = 1.5 GHz) = 4.7 radians = 1.56 radians (modulo 2$\pi$), if it were electromagnetic pickup. The actual phase shift (modulo 2$\pi$) is about 3 radians, which does not match $\phi$. Hence, it is most likely not due to direct electromagnetic pickup. In this case, we could not test the control sample and hence cannot confirm this independently.

\begin{figure}[!h]
\centering
\includegraphics[width=4in,angle=270]{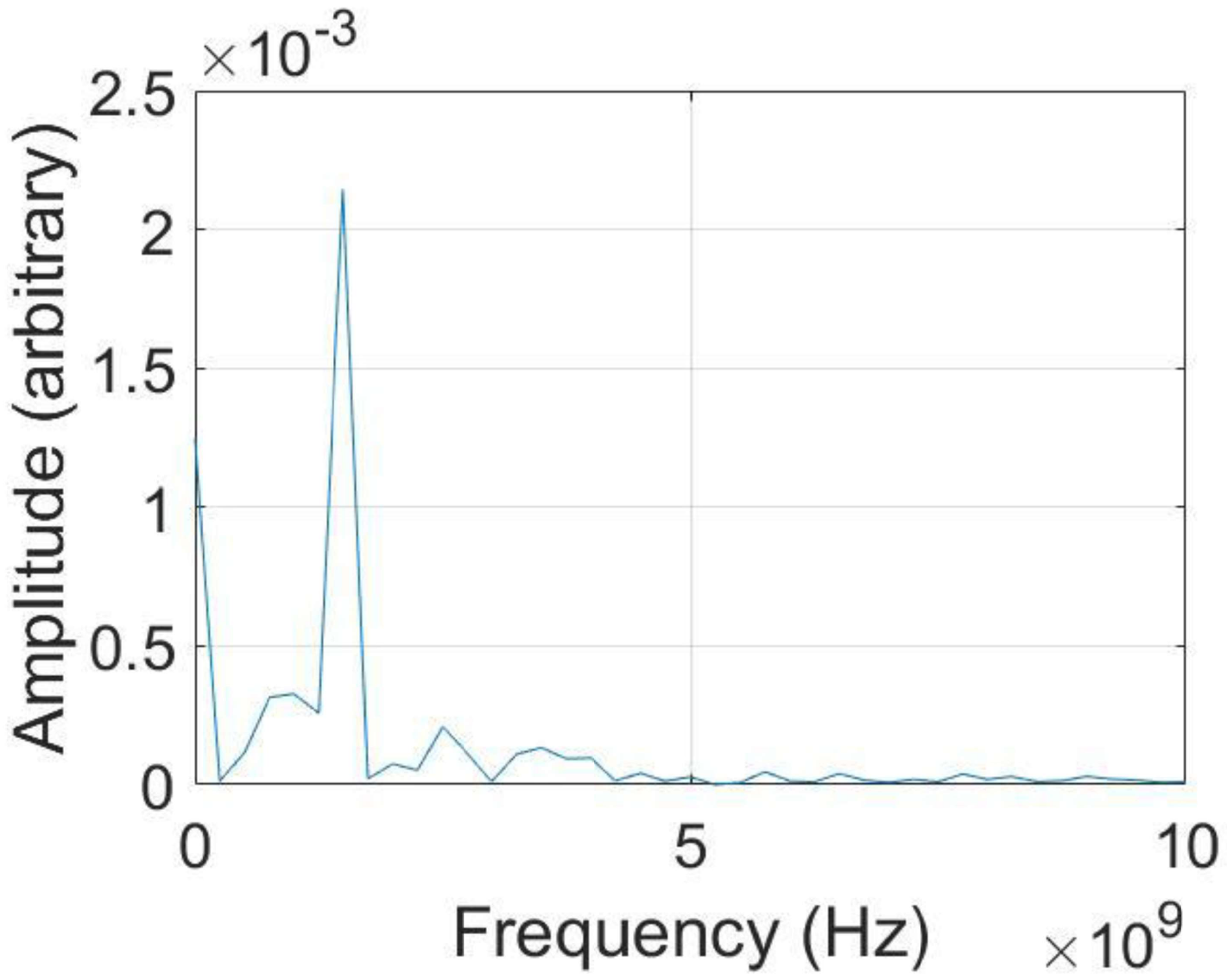}
\caption{Fast Fourier transform of the received signal at 1.5 GHz excitation.}
\label{fig:1.5FFT}
\end{figure}

In Fig. \ref{fig:1.5FFT} , we show the Fourier transform of the received signal. Note that there are frequency components in the received signal at 750 MHz and 2.5 GHz which also are present in the output when the input signal frequency is 2.4 GHz. Since these are independent of the input signal frequency, they must correspond to intrinsic modes of the system.

A major difference between EM excitations at 1.5 GHz and 2.4 GHz is that in the former case, the EM wave excites both intrinsic and extrinsic spin wave modes in the nanomagnets and hence the output voltage has frequency components at the frequencies of the intrinsic modes (750 MHz and 2.5 GHz) as well as the frequency of the EM wave (1.5 GHz). The latter is the frequency of the extrinsic mode spawned by the excitation. In fact, the signature of the extrinsic mode is dominant over those of the intrinsic modes in the output voltage. In contrast, the 2.4 GHz EM wave does not produce a 2.4 GHz frequency component in the output voltage but instead produces components at 750 MH and 2.5 GHz which are associated with intrinsic modes of the nanomagnet array. However, this does not mean that the 2.4 GHz excitation does not produce any extrinsic mode. It may or may not have produced one at 2.4 GHz, but if it did, it may not be resolvable from the intrinsic mode at 2.5 GHz since these two frequencies are very close. But in the context of the receiving antenna, these discussions are somewhat academic. What matters is that the incident EM signal produces a voltage output, which is all that is necessary to implement the receiving function.

\section*{Receiver gain at 1.5 GHz}

We can use Equation (1) from the main paper to calculate the receiver gain at 1.5 GHz:
\vspace{-0.1in}
\begin{equation*}
   \frac{P_r}{P_t} =\frac{V_{out}^2}{V_{in}^2} \approx (30)^2 = \frac{G_t G_r  \lambda^2}{4\pi R^2} 
\end{equation*}

Using $G_t$ = 9 dB = 8, $\lambda$ = 20 cm, $R$ = 6 in = 15.24 cm, we get $G_r$ = 0.07 = -11.5 dB.

\end{document}